\documentclass[aps,pra,twocolumn,groupedaddress,amsmath,amssymb,10pt]{revtex4}
\usepackage{epsfig}
\usepackage{amssymb}
\usepackage{amsmath}
\usepackage{graphicx}
\usepackage{amsfonts}
\usepackage{revsymb}
\usepackage{bm}
\usepackage{amsmath}
\usepackage{amssymb}
\usepackage{latexsym}
\usepackage{hyperref}
\usepackage{cleveref}

\begin{document}

\title{Ginzburg-Landau approximation for the Hubbard model
in the external magnetic field }
\author{L. B. Dubovskii}
\author{S. N. Burmistrov}
\affiliation{NRC "Kurchatov Institute", 123182 Moscow, Russia}

\begin{abstract}
The Hubbard model is studied in the external magnetic field. The analysis is carried out phenomenologically within the framework of the Ginzburg-Landau theory with the order parameter describing the opposite spin electrons. The study is performed for the nearly half-filled lower Hubbard band in the metallic state. The final equations are the Pauli-like ones for the opposite spins and nonlinear as a result of interaction between electrons with the opposite spins. The equations can analytically be solved for the spatially homogeneous distributions in a number of most interesting cases. In particular, the problem on the metal-insulator transition is analyzed for the nearly half-filled Hubbard sub-bands. The critical magnetic field at which the transition from the metallic state to the insulator one takes place is found under the paramagnetic spin effect.
\end{abstract}
\maketitle

\section{Introduction}
The traditional electronic theory of metals starts from the weakly interacting electrons moving in the periodic crystal field and is described by the standard band conception \cite{brikos}. According to the band theory, various transition metal oxides should be conductors since they have an odd number of electrons per unit cell. The typical system of this kind is nickel oxide NiO. However, it is found that such compounds behave like insulators in reality. Nevill Mott \cite{Mott49} predicted that the anomaly could be explained by involving the Coulomb interaction between electrons and proposed the model for NiO as an example of an insulator for transition metal oxides. Such anomalous state is referred to as the Mott insulator. The Mott insulator state emerges provided that the repulsive Coulomb potential is sufficiently large in order to produce the gap in the electron energy spectrum.
\par
The simplest approach to this problem is to apply the Hubbard model  \cite{HubbardI}. Describing the metal system, J.~Hubbard starts from the usual electron-ion Hamiltonian for a hypothetical partly-filled narrow \textit{s}-band containing $n$ electrons per atom. In the case of narrow energy bands J.~Hubbard has emphasized that one should take the atomicity of the electron distribution into account and  employ a very simple approximate representation of the electron-electron interactions. In fact, this approximation is mathematically much simpler to handle than the Coulomb interaction in itself. 
\par 
On the whole, the analysis leads to the simplified Hamiltonian as 
\begin{equation}\label{hub1}
\hat{H}=\sum_{i,j,\sigma}T_{ij}c_{i\sigma} ^{+}c_{j\sigma}+\frac{1}{2}I\sum_{i,\sigma}n_{i,\sigma}n_{i,-\sigma}-
I\sum_{i,\sigma}\nu_{ii}n_{i,\sigma}
\end{equation}
where $n_{i,\sigma}=c_{i\sigma} ^{+}c_{i\sigma}$. Here $c_{\bm{k}\sigma}$ and $c_{\bm{k}\sigma} ^+$ are the annihilation and creation operators of electrons with spin $\sigma$ in the Bloch state and the transition amplitude reads 
$$
T_{ij}=N^{-1}\sum _{\bm{k}}\epsilon _{\bm{k}}\exp{i\bm{k} (\bm{R}_i -\bm{R}_j)},\quad \nu _{ii}=N^{-1}\sum _{\bm{k}}\nu _{\bm{k}}\, ,
$$
$N$ being the number of atoms. The sum in \eqref{hub1} runs over all the atomic sites $\bm{R}_i$. The magnitude $I$ is the Coulomb repulsion of the electrons at the same transition oxide atom.  The quantity  
$\epsilon _{\bm{k}}$ is the electron band spectrum and magnitude $\nu _{\bm{k}}$ is the occupation  number of the electron states in the band.
\par
It should be emphasized that the interactions in the Hamiltonian $\hat{H}$  are between the opposite-spin electrons alone and these interactions are completely local. The physical reason for the completely on-site  interactions in $\hat{H}$ (\ref{hub1}) is determined by the fact that the electron interaction is only significant for the electrons belonging to the same transition atom. The electron interaction from various transition atoms is negligible. So, despite the band motion of \textit{d}-electrons, the electrons at the same atom are strongly correlated with each other and correlated weakly with the electrons from other atoms. Such intra-atomic correlations are responsible for the physical properties  as compared with the system of isolated atoms.
\par
The most significant result from this Hamiltonian  \cite{HubbardI} is that the electron conduction band, arising in the conventional electron theory of metals at half-filling, proves to be split into two sub-bands. The lower sub-band is completely filled with the conduction electrons. The upper sub-band proves to be completely empty.
In the Hubbard electron conduction band  there are electrons of the spin-up and spin-down directions with the purely local interaction.
\par
The purpose of our paper is a construction of phenomenological approach for the Hubbard model in the Ginzburg-Landau approximation  and application it for most interesting problems. The behavior in the magnetic field is significant for the metal-insulator transitions in the first turn. This is interesting both for the paramagnetic case and for the case of the orbital electron motion.  The static spin structures and spin wave problems can be analyzed within the framework of this approach as well.

\section{Hubbard model in the Ginzburg-Landau approach}

The description of the Hubbard model in the Ginzburg-Landau approach can be performed by introducing the order parameter with components $u_{\uparrow}(\bm{r})$ and  $u_{\downarrow}(\bm{r})$. The component $u_{\uparrow}(\bm{r})$ describes a spin-up electron  and component  $u_{\downarrow}(\bm{r})$ does a  spin-down one. The interactions in the Hamiltonian $\hat{H}$ \eqref{hub1}  are between the opposite-spin electrons alone. We introduce  the same in the Ginzburg-Landau approximation.
\par
In zero magnetic field the components $u_{\uparrow}(\bm{r})$ and  $u_{\downarrow}(\bm{r})$ are real quantities. However, the components become the complex quantities in the finite magnetic field. The construction of the Ginzburg-Landau functional $\Phi$ can be performed as usual (cf. \cite{DUB,LandLif2, LBD})
\begin{gather}
\Phi=\Phi [u_{\uparrow}(\bm{r}), u_{\downarrow}(\bm{r})]=\int d\bm{r}\,\emph{F}(u_{\uparrow}(\bm{r}), u_{\downarrow}(\bm{r}))\, , \nonumber
\\
\emph{F}(u_{\uparrow}(\bm{r}), u_{\downarrow}(\bm{r}) )=\alpha\sum _{\sigma} \left({u_{\sigma}^*(\bm{r}) u_{\sigma}(\bm{r})}\right)\quad\quad\quad \nonumber
\\ 
+\frac{1}{2}\beta \sum _{\sigma} \left({u_{\sigma}^*(\bm{r}) u_{\sigma}(\bm{r})}\right)\left({u_{-\sigma}^*(\bm{r}) u_{-\sigma}(\bm{r})}\right) \nonumber 
\\ 
+\frac{\hbar ^2}{2m}\sum\limits_{\sigma}
\left(\nabla u_{\sigma}^*(\bm{r})\nabla u_{\sigma} (\bm{r})\right) .
\label{glrn1}
\end{gather}
The variable $\alpha$ characterizes the electron spectrum with the both spins $\sigma =\uparrow ,\, \downarrow$. The magnitude $\beta$ determines the electron interaction between the opposite spins  $\sigma =\uparrow ,\, \downarrow$.
\par
Varying the functional \eqref{glrn1} leads to the following equation:
\begin{gather}
-i\hbar\frac{\partial}{\partial t}u_{\sigma} (\bm{r},t)=-\frac{\hbar ^2}{2m}\nabla^2u_{\sigma}(\bm{r},t)+\alpha  u_{\sigma} (\bm{r},t) \nonumber 
\\ 
+\beta\left({u_{-\sigma}^{*}(\bm{r},t)}u_{-\sigma}(\bm{r},t)\right)u_{\sigma}(\bm{r},t), \quad \sigma =\uparrow ,\, \downarrow .
\label{hub13711}
\end{gather}
Equation \eqref{hub13711} allows us to determine the excitation spectrum in the electronic system.
As a first step, we consider only the static and homogeneous coordinate-independent electron density   $\rho _{\sigma}$ in the metal for a fixed spin. In this case the magnitude $\rho _{\sigma}(\bm{r},t)= \left({u_{\sigma}^{*}(\bm{r},t)}u_{\sigma}(\bm{r},t)\right) \equiv n_{\sigma}(0)$ is independent of coordinate $\bm{r}$ and time $t$. The two equations for different spins $(\ref{hub13711})$   represent  the linear system for $u_{\sigma}(\bm{r},t)$ with $\sigma =\uparrow ,\, \downarrow $
\begin{gather}
-i\hbar\frac{\partial}{\partial t}u_{\sigma} (\bm{r},t)=-\frac{\hbar ^2}{2m}\nabla^2u_{\sigma}(\bm{r},t)+\alpha  u_{\sigma} (\bm{r},t) \nonumber
\\ 
+\beta\left({u_{-\sigma}^{*}(\bm{r},t)}u_{-\sigma}(\bm{r},t)\right)u_{\sigma}(\bm{r},t) , \quad \sigma =\uparrow ,\, \downarrow .
\label{hub13721}
\end{gather}
The order parameter $u_{\sigma}(\bm{r})$ corresponds fully to the description of the Hubbard model (\ref{hub1}) with the interactions between the opposite-spin electrons. In addition, the electron interactions are completely local as in the Hubbard model.
\par
For the case of homogeneous coordinate-independent density of electrons, the system in Eq.~\eqref{hub13711} reduces to
\begin{gather}
\alpha u_{\sigma} + \beta n_{-\sigma}(0)u_{\sigma}=0 , \quad\sigma =\uparrow ,\, \downarrow , \nonumber 
\\
n_{\uparrow}(0)=n_{\downarrow}(0) =-\alpha /\beta  .
\label{hub13723}
\end{gather}
The value $\beta <0$ refers to the Coulomb repulsion. If $\alpha\gg\mid \beta \mid$, the magnitude $n_{\sigma}(0)$ can arbitrarily be large and is limited by the electron Fermi statistics alone. Correspondingly, no more than two electrons with the opposite spins can be in the conduction band, i.e.
$$
n_{\sigma}(0)+n_{-\sigma}(0)\leqslant 2 , \quad \sigma =\uparrow ,\, \downarrow.
$$
For $\mid\beta\mid\gg\alpha$, we have $n_{\sigma}(0)\ll 1 $. The system \eqref{hub13721} can exactly be solved  using the Fourier transformation. We can seek for the solution as
$$
u_{\sigma}(\bm{r},t)=u_{\sigma\varepsilon}(\bm{k})\exp\biggl[\frac{i}{\hbar}(\bm{k}\bm{r}+\varepsilon t)\biggr] .
$$
As a result, we arrive at two connected linear equations with $\sigma =\uparrow ,\, \downarrow$
$$
\varepsilon u_{\sigma\varepsilon}(\bm{k})=\frac{k^2 }{2m}u_{\sigma\varepsilon}(\bm{k})+\alpha  u_{\sigma\varepsilon}(\bm{k}) +\beta n_{-\sigma}(0)u_{\sigma\varepsilon}(\bm{k}) .
$$ 
 So, we obtain two different electronic energy branches for various spins. At the same time these energy branches are connected with each other via electron densities
$$
\varepsilon _{\sigma}=\frac{k^2 }{2m}+\alpha _{\sigma} +\beta n_{-\sigma}(0), \quad\sigma =\uparrow ,\, \downarrow .
$$
\par
Introducing the magnetic field $\bm{h}(\bm{r})$ and vector potential $\bm{A}(\bm{r})$ into the system is performed by the conventional way \cite{LandLif2,LBD} similar to that in superconductor. So, we write the magnetic field-dependent part of the total functional as  
\begin{gather}\label{hub12}
\emph{F}_h =\frac{\bm{h}^2}{8\pi}+\frac{\hbar ^2}{2m}\sum _{\sigma}\left|\biggl(\nabla-\frac{ie}{\hbar c}\bm{A}\biggr)u_{\sigma}(\bm{r})\right|^2 ,
\\
\bm{h} =\text{curl}\,\bm{A},\quad \text{div}\,\bm{h}=0,\quad \text{curl}\,\bm{h}=\frac{4\pi}{c}\bm{j}(\bm{r}), \nonumber
\\
\bm{j}(\bm{r})=\sum _{\sigma}\biggr(\frac{ie\hbar}{2m}\bigl[\bigl(\nabla u_{\sigma}^*(\bm{r})\bigr) u_{\sigma}(\bm{r})-u_{\sigma}^*(\bm{r}) \nabla u_{\sigma}(\bm{r})\bigr] \nonumber
\\ 
-\frac{e^2}{mc}\bm{A}u_{\sigma}^*(\bm{r})u_{\sigma}(\bm{r}) +\mu\,  \text{curl}\,\bigl[ u_{\sigma}^*(\bm{r}) \bm{\sigma} u_{\sigma}(\bm{r})\bigr]\biggr), \nonumber
\end{gather}
$\mu$ being the effective Bohr magneton. In the magnetic field the order parameter  $u_{\sigma}(\bm{r})$ satisfies the following equation:
\begin{multline}\label{hub13}
-\frac{\hbar ^2}{2m}\biggl(\nabla-\frac{ie}{\hbar c}\bm{A}(\bm{r})\biggr)^2 u_{\sigma}(\bm{r})+\alpha u_{\sigma}(\bm{r})  
\\ 
+\beta\vert u_{-\sigma}(\bm{r},t)\vert ^2u_{\sigma}(\bm{r}) +\mu\bigl(\bm{\sigma}\bm{h}(\bm{r})\bigr)u_{\sigma}(\bm{r}) =0  
\end{multline}
where $\sigma =\uparrow ,\, \downarrow$. This system of equations~\eqref{hub13} is nonlinear.
Provided that the quantity $\vert u_{-\sigma}(\bm{r},t)\vert^2\equiv \text{const}$ and is spatially homogeneous, nonlinear equation~(\ref{hub13}) becomes linear and can exactly be solved. In this case  the quantity $\vert u_{-\sigma}(\bm{r},t)\vert ^2\equiv n_{-\sigma}(0)$ represents the number of electrons in the cell with  $\sigma =\downarrow ,\, \uparrow  $ and does not depend on the coordinate $\bm{r}$.
\par
The term  with vector potential $\bm{A}(\bm{r})$  in Eq.~\eqref{hub13} describes the orbital motion of electrons in the metal and the term $\mu\bm{h}(\bm{r})$ represents the Pauli paramagnetism.

\section{Metal-insulator transition}

One of most significant results of analyzing  Hamiltonian $\hat{H}$ (\ref{hub1}) is that the conduction bands, arising in the traditional electronic theory of metals at half-filling, turn out to be split into two sub-bands (Hubbard sub-bands, see \cite{HubbardI, HubbardII, HubbardIII, HubbardIV, HubbardV, HubbardVI, HubbardS, Zaitsev76, Moskalenko, The Hubbard Model, Alexei, Niu1, Niu2}).  In this case the lower sub-band proves to be completely filled with the conduction electrons and the upper one proves to be completely empty. The energy gap appears between these two sub-bands. It is necessary to emphasize that the Coulomb interaction should be sufficiently strong in this scenario. The system becomes an insulator. This is the physical reason for the Mott insulator. The transition metal oxides are  the brilliant examples of Mott insulator.
\par
If $\alpha =\mid\beta\mid$, we have according to Eq.~\eqref{hub13723}
\begin{equation}\label{hub1372311a}
n_{\uparrow }(0)=n_{\downarrow}(0) =-\alpha /\beta =1.
\end{equation}
So, the lower Hubbard subband will completely be filled. For the almost filled Hubbard band, this means that the phenomenological constants in the Ginzburg-Landau equations are connected by the following relation:
\begin{equation}\label{hub20}
\vert\beta\vert =\frac{\alpha}{1-\delta} .
\end{equation}
The small constant $\delta$ shows how much the filling in the lower Hubbard band differs from the full filling. The complete filling  corresponds to an insulator.  At $\delta =0$, the metal goes over to the insulator state.
\par
We will consider simplest problem when the Pauli paramagnetism near the half-filling of the electron band transfers the metal state  to the insulator one  under influence of external magnetic field.
The case of small $\delta \ll 1$ will be considered in the following magnetic field:
$$
\bm{h}=\bm{i} h_x +\bm{k} h_z(x),
 $$
$\bm{i}$ and $\bm{k}$ being unit vectors in the coordinate directions.  Let magnetic field vary in the metal in the \textit{x} direction. The axis \textit{z} is normal to the the metal surface. The axis \textit{x} is parallel to the metal surface. The appropriate magnitude of the vector potential, directed along the \textit{y} axis, reads
$$
\bm{A}=\bm{j}\left(\int _{-\infty} ^x  h_z(x')dx'-zh_x\right).
$$
Here component $ h_z$ is a function of $x$ and the component $h_x$ is independent of $x$ since the Maxwell equation $\text{div}\,\bm{h} =0$ should be satisfied.
\par
For simplicity, we assume that, due to scattering in the system, the orbital motion of electrons is suppressed in the metal and we neglect the vector potential $\bm{A}$.
 Then equations \eqref{hub13} take the form
\begin{multline}
-\frac{\hbar ^2}{2m}\nabla ^2 u_{\sigma}(\bm{r})+\alpha u_{\sigma}(\bm{r}) +\beta n_{-\sigma}(0)
u_{\sigma}(\bm{r}) 
\\ 
+\mu \bigl(\bm{\sigma}\bm{h}(\bm{r}) \bigr)u_{\sigma}(\bm{r}) =0,\quad\sigma =\uparrow ,\, \downarrow .
\label{hub1312a}
\end{multline}
Here we have replaced the magnitude $\vert u_{-\sigma}\vert ^2$ with $n_{-\sigma}(0)$ as we consider the  static and homogeneous coordinate-independent density of electrons in the metal.
\par
We use the following notation:
$$
(\bm{\sigma}\bm{h})=
\begin{pmatrix}
h_z\,\,\,\, h_x
\\
h_x\,\,\, -h_z
\end{pmatrix}.
$$
The homogeneous and coordinate-independent solution for $u_{\sigma}(\bm{r})\,\,\,(\sigma =\uparrow ,\, \downarrow)$ can be written as the following system of equations:
\begin{gather}
\alpha u_{\uparrow}+\beta n_{\downarrow}(0)u_{\uparrow}+\mu (h_z u_{\uparrow}+h_xu_{\downarrow})=0,
\nonumber
\\ 
\alpha u_{\downarrow}+\beta n_{\uparrow}(0)u_{\downarrow}+\mu (h_x u_{\uparrow}-h_z u_{\downarrow})=0\, .
\label{hub141}
\end{gather}
Let us put $u_{\uparrow}\neq 0$ and $u_{\downarrow}\neq 0$. Then from this equation we arrive at the following one:
\begin{equation}\label{hub15}
\alpha +\beta n_{\downarrow}(0)+\mu h_z +\frac{(\mu h_x)^2}{\alpha +\beta n_{\uparrow}(0)-\mu h_z } =0 \, .
\end{equation}
If the number of electrons with the both spin directions equals each other, i.e.  $n_{\downarrow}(0)=n_{\uparrow}(0)=n$, we have
\begin{equation}\label{hub151}
n=\frac{\alpha}{\mid\beta\mid}\pm\frac{\mu}{\mid\beta\mid}\sqrt{(h_z ^2-h_x ^2)}\, .
\end{equation}
Here we use that  $\beta$ is negative $(\beta <0)$ for the Coulomb interaction between two electrons.
Then we can write Eq.~(\ref{hub151}) for $h_x=0$ in the simplified form
\begin{equation}\label{hub152}
n=\frac{\alpha}{\mid\beta\mid}+\frac{\mu}{\mid\beta\mid}h_z \, .
\end{equation}
If we put the magnetic field as
$$
h_z=\alpha \frac{\delta}{\mu}\, ,
$$
the magnitude $n$ becomes unity. Then the system goes over from the metallic state to insulator one. The critical magnetic field reads
\begin{equation}\label{hub16a}
h_{cr}=\frac{\alpha\delta}{|\mu|}
\end{equation}
and means that the lower Hubbard band becomes completely filled. The same situation realizes as well  when  certain number of electrons is in the upper Hubbard sub-band.  We can transfer the system to the insulator state by appropriate selection of the direction and magnitude of the magnetic field in \eqref{hub16a}.
\par
In the case when the electron densities with the different spin projections are not equal to each other,  such system becomes ferromagnetic according to \cite{HubbardI}. By governing the magnetic field according to Eq.~\eqref{hub15}, it is possible to control the states of the ferromagnetic system.

\section{Conclusion}

To conclude, we have analyzed the Hubbard model~\cite{HubbardI} in the external magnetic field within the Ginzburg-Landau approximation. The starting point is the introduction of two-component order parameter with components $u_{\uparrow}(\bm{r})$ and $u_{\downarrow}(\bm{r})$. The component $u_{\uparrow}(\bm{r})$ describes an electron with the spin-up orientation and the component $u_{\downarrow}(\bm{r})$ refers to an electron with spin-down. The electron interactions are essential only for the electrons at the same transition atom. The electron interactions at various transition atoms are neglected. This implies that the electron interactions are fully local and effective only between the electrons with the opposite spins. The external magnetic field mediates an additional interaction between electrons of different spin directions even for the homogeneous system, see  \eqref{hub1312a} and \eqref{hub15}. The strong Coulomb repulsion results in the low density of electrons in the conduction band.
\par
The resulting equations \eqref{hub13}, which are the Pauli-like ones for the opposite spins and nonlinear due to  interaction of electrons with the different spins, can analytically be solved in a number of important cases. For example, the problem of metal-insulator transition in the external magnetic field is exactly solved for the case of  the nearly half-filled  lower Hubbard sub-band. The critical magnetic field  $h_{cr}\sim\alpha\delta /|\mu|$ \eqref{hub16a} is governed by the parameters of  electron spectrum and Coulomb repulsion. Provided that the electron densities with the various spin projections are different, the system becomes ferromagnetic. Varying the magnetic field makes it possible to control the ferromagnetic behavior in the system.

\end{document}